\documentclass[reprint,aps,prd,nofootinbib,a4paper,10pt,twocolumn,superscriptaddress]{revtex4-2}
\usepackage{hyperref} 
\usepackage{changes}
\usepackage[section]{placeins}
\usepackage[titletoc]{appendix}
\usepackage{xcolor}
\usepackage{dsfont}
\usepackage{bm}
\usepackage{mathtools} 
\usepackage{amsfonts}
\usepackage{enumerate}
\usepackage{graphicx,subfigure}
\DeclareMathAlphabet{\mathpzc}{OT1}{pzc}{m}{it}

\newcommand{\sayy}[1]{`#1'}
\providecommand{\href}[2]{#2}

\def\be{\begin{equation}}
\def\ee{\end{equation}}
\def\bea{\begin{eqnarray}}
\def\eea{\end{eqnarray}}

\def\sig{\sigma}

\def\la{\langle}
\def\ra{\rangle}
\def\Eu{ \mathfrak{H} }

\def\obs{\mathcal{O}}
\def\emi{\mathcal{E}}
 
\def\d{\mathrm{d}}
\definecolor{MyB}{rgb}{0.1,0.1,1.0}

\definecolor{mygreen}{rgb}{0,0.5,0}

\begin{document} 

\title{Redshift drift in a universe with structure I: Lema\^itre-Tolman-Bondi structures with arbitrary angle of entry of light}  

\author{Sofie Marie Koksbang} 
\email{koksbang@cp3.sdu.dk}
\affiliation{CP$^3$-Origins, University of Southern Denmark, Campusvej 55, DK-5230 Odense M, Denmark}

\author{Asta~Heinesen}
\email{asta.heinesen@ens--lyon.fr}
\affiliation{Univ Lyon, Ens de Lyon, Univ Lyon1, CNRS, Centre de Recherche Astrophysique de Lyon UMR5574, F--69007, Lyon, France}

\begin{abstract} 
We consider the redshift drift and position drift associated with astrophysical sources in a formalism that is suitable for describing emitters and observers of light in an arbitrary spacetime geometry, while identifying emitters of a given null-geodesic bundle that arrives at the observer worldline.
We then restrict the situation to the special case of a Lema\^itre-Tolman-Bondi (LTB) geometrical structure, and solve for lightrays propagating through the structure with arbitrary impact parameters, i.e., with arbitrary angles of entry into the LTB structure. 
The redshift drift signal emitted by comoving sources and viewed by a comoving observer turns out to be dominated by Ricci curvature and electric Weyl curvature contributions as integrated along the connecting 
light ray. 
This property simplifies the computations of the redshift drift signal tremendously, and we expect that the property extends to more complicated models including Swiss-cheese models. 
When considering several null rays with random impact parameters,  
the mean redshift drift signal is well approximated by a single Ricci focusing term. This suggests that the measurement of cosmological redshift drift can be used as a direct probe of the strong energy condition in a realistic universe where photons pass through many successive structures.
\end{abstract}
\keywords{Redshift drift, relativistic cosmology, observational cosmology} 

\maketitle

\section{Introduction}
Redshift drift is the temporal change in redshift  of light arriving from a distant source as viewed by the observer  \cite{1962ApJ...136..319S,1962ApJ...136..334M}. The detection of redshift drift is a cornerstone of upcoming precise cosmological measurements \cite{Liske:2008ph}, and makes possible the direct determination of kinematic properties of the Universe, which would otherwise rely on indirect inference and the assumption of a cosmological model. 
Redshift drift is a probe of dark energy within the Friedmann-Lema\^{\i}tre-Robertson-Walker (FLRW) universe models \cite{Loeb:1998bu,Lobo:2020hcz}, but might be used as a probe of violation of the strong energy condition within much broader universe geometries  \cite{Heinesen:2021nrc,Koksbang:2019glb}. Redshift drift might also be used as a test of the FLRW spacetime conjecture  \cite{Heinesen:2020pms,Koksbang:2021qqc}. 

The redshift drift signal has mostly been analysed within the FLRW universe models, but analytical and numerical investigations have  also been carried out within Stephani, Lema\^{\i}tre-Tolman-Bondi (LTB), Bianchi I, and Szekeres models \cite{Uzan:2008qp,2011PhRvD..83d3527Y,Mishra:2012vi,Balcerzak:2012bv,Fleury:2014rea,Koksbang:2019glb,Koksbang:2020zej,Koksbang:2015ctu,Mishra:2014vga,Balcerzak:2012mka}. Convenient representations of redshift drift within arbitrary spacetime geometries have recently been formulated \cite{Korzynski:2017nas,Heinesen:2020pms}, and a promising numerical tool for fast computation of drift effects 
for a given specified metric description has been proposed \cite{Grasso:2021iwq,Grasso:2021zra}. 

The redshift drift representation for an arbitrary geometrical setting as formulated in \cite{Heinesen:2020pms}  
is useful for analysing potential systematic departures from the FLRW redshift drift prediction induced by local structures  \cite{Heinesen:2020pms}, for independent observational tests of the strong energy condition \cite{Heinesen:2021nrc}, and for performing model-independent cosmographic analyses of data \cite{Heinesen:2021qnl}.   
The representation is furthermore useful for investigating  individual curvature and kinematic contributions to the final redshift drift signal within model-universes of interest. 

In this paper,  
we consider the redshift drift in a class of LTB models with light propagating through the structure with arbitrary angles of entry. The investigated LTB model profile describes a central underdensity surrounded by a steep overdensity, and thus might be used as a crude model of a void with surrounding filaments of galaxies.
Using the framework developed in \cite{Heinesen:2020pms,Heinesen:2021nrc,Heinesen:2021qnl} to decompose the redshift drift signal allows us to analyse the hierachy of multipole terms contributing to the signal along the light beam, when the light passes through the LTB structure.  
We analyse the relative magnitudes of the individual terms as well as cancellation effects relating to these terms;  in particular, we analyse the conjecture that Ricci focusing dominates the redshift drift signal when light rays are traversing many structures by considering the situation where many light rays traverse a single structure with different (random) impact parameters.

In section~\ref{sec:dec} we review the general expression for the redshift drift signal in terms of the physically interpretable multipole decomposition, and consider the multipole coefficients in the special case of an LTB spacetime. 
In section~\ref{sec:num} we describe the details of our analysis regarding the LTB model parameterization and light propagation.  
In section~\ref{sec:results} we describe the results of our analysis, and we conclude in section~\ref{sec:conclusion}.

\vspace{5pt} 
\noindent
\underbar{Notation and conventions:}
Units are used in which $c=1$. Greek letters $\mu, \nu, \ldots$ label spacetime
indices in a general basis while Latin letters $i,j,\ldots$ denote spatial indices relative to a specified foliation frame. Einstein notation is used such that repeated indices are summed over.  
The signature of the spacetime metric $g_{\mu \nu}$ is $(- + + +)$ and the connection $\nabla_\mu$ is the Levi-Civita connection. 
Round brackets $(\, )$ containing indices denote symmetrisation in the involved indices and square brackets $[\, ]$ denote anti-symmetrisation. 
Bold notation $\bm V$ for the basis-free representation of vectors $V^\mu$ is used occasionally. A subscripted comma followed by an index indicates partial derivative. 


\section{Multipole decomposition of the redshift drift signal} 
\label{sec:dec}
In this section we consider the multipole decomposition of the redshift drift signal developed in \cite{Heinesen:2020pms,Heinesen:2021nrc,Heinesen:2021qnl}, which is appropriate for analysing kinematic and curvature 
contributions to the drift of the redshift of a source. 
In section~\ref{sec:general} we consider the decomposition in a general spacetime setting with an arbitrary observer congruence, and we then move on to analyse the special case of an LTB metric with comoving observers in section~\ref{sec:LTB}. 

\subsection{General spacetime}
\label{sec:general} 
Following \cite{Heinesen:2020pms,Heinesen:2021nrc,Heinesen:2021qnl} we consider a general congruence of emitters and observers (denoted the \sayy{observer congruence}) in an arbitrary spacetime. We let the observer congruence be generated by the  4-velocity field $\bm u$, and parameterized by the proper time function $\tau$ satisfying $\dot{\tau} = 1, $where $\dot{} \equiv u^\mu \nabla_\mu$ is the directional derivative along the observer congruence flow lines. The general kinematic decomposition associated with the frame of the observer congruence is 
\bea 
\label{def:expu}
&& \nabla_{\nu}u_\mu  = \frac{1}{3}\theta h_{\mu \nu }+\sig_{\mu \nu} + \omega_{\mu \nu} - u_\nu a_\mu  \ , \nonumber \\ 
&& \theta \equiv \nabla_{\mu}u^{\mu} \, ,  \quad \sig_{\mu \nu} \equiv h_{ \la \nu  }^{\, \beta}  h_{  \mu \ra }^{\, \alpha } \nabla_{ \beta }u_{\alpha  }  \, , \nonumber \\ 
&& \omega_{\mu \nu} \equiv h_{  \nu  }^{\, \beta}  h_{  \mu }^{\, \alpha }\nabla_{  [ \beta}u_{\alpha ] }   \, , \quad  a^\mu \equiv \dot{u}^\mu \,  , 
\eea 
where $h_{ \mu }^{\; \nu } \equiv u_{ \mu } u^{\nu } + g_{ \mu }^{\; \nu }$ is the spatial projection tensor relative to the observer congruence, and where $\la \ra$ is the traceless and symmetric part of a spatially projected tensor\footnote{See  \cite{Spencer:1970} for details on the unique traceless decomposition of spatial symmetric tensors, and see \cite{Heinesen:2020bej} for the explicit decomposition for tensors with up to six indices.}.  

We may consider two causally connected members of the observer congruence with worldlines $\gamma_o$ and $\gamma_e$ passing through the events of observation $\obs$ and emission $\emi$ of a null geodesic ray. 
Let $\bm k$ be the 4-momentum of a 4-dimensional non-caustic geodesic null congruence that contains this null ray, and which creates a bijection between $\gamma_o$ and $\gamma_e$ in a neighbourhood around the points $\obs$ and $\emi$. 
We define the photon energy as measured by members of the observer congruence $E \equiv - u^\mu k_\mu$,
and the spatial unit-vector $e^\mu \equiv u^\mu - \frac{1}{E} k^\mu$ describing the direction of observation or the \sayy{viewing angle} of the light ray as seen by the same observers. 
We introduce the drift of the viewing angle
\bea
\label{positiondrift}
\kappa^\mu \equiv  h^{ \mu }_{\; \nu }   \dot{e}^\nu \, , 
\eea  
which describes the change of spatial direction of incoming light as seen in the observer congruence reference frame. 
When \eqref{positiondrift} is evaluated at $\obs$, it represents the position drift of the astrophysical emitter as viewed on the observer's sky.

The drift of the redshift, $z \equiv E_{\gamma_e}/E_{\gamma_o} - 1$ as observed by the observer along $\gamma_o$ in the vicinity of $\obs$ can be written as the integral 
\bea
\label{redshiftdriftint}
\frac{d z}{d \tau} \Bigr\rvert_{\obs} = E_\emi \! \! \int_{\lambda_\emi}^{\lambda_\obs} \! \! d \lambda \,   \Pi     \, , \qquad z \equiv \frac{E_\emi}{E_\obs} - 1
\eea  
where $\lambda$ is an affine parameter along the null geodesic congruence satisfying $k^\mu \nabla_\mu \lambda = 1$.  Using the traceless multipole decomposition in $\bm{e}$ and $\bm{\kappa}$, the integrand, $\Pi$, can be written as \cite{Heinesen:2021qnl}
\bea
\label{Pimultikappa}
\hspace*{-0.65cm} \Pi &=&  -  \kappa^\mu \kappa_\mu  + \Sigma^{\it{o}}    +  e^\mu \Sigma^{\bm{e}}_\mu    +       e^\mu   e^\nu \Sigma^{\bm{ee}}_{\mu \nu} + e^\mu   \kappa^\nu \Sigma^{\bm{e\kappa}}_{\mu \nu}    
\eea  
with coefficients 
\bea
\label{Picoefkappa}
&& \Sigma^{\it{o}} \equiv  - \frac{1}{3} u^\mu u^\nu R_{\mu \nu}     + \frac{1}{3}D_{\mu} a^{\mu} + \frac{1}{3} a^\mu a_\mu    \, , \nonumber  \\   
&&  \Sigma^{\bm{e}}_\mu  \equiv   - \frac{1}{3}  \theta a_\mu    -   a^{ \nu} \sigma_{\mu \nu}  + 3 a^{ \nu} \omega_{\mu \nu}    - h^{\nu}_{\mu} \dot{a}_\nu \, ,  \nonumber  \\   
&& \Sigma^{\bm{ee}}_{\mu \nu} \equiv     a_{\la \mu}a_{\nu \ra }  + D_{  \la \mu} a_{\nu \ra }    -  u^\rho u^\sigma  C_{\rho \mu \sigma \nu}   -  \frac{1}{2} h^{\alpha}_{\, \la \mu} h^{\beta}_{\, \nu \ra}  R_{ \alpha \beta }  \,   , \nonumber  \\   
&& \Sigma^{\bm{e\kappa}}_{\mu \nu} \equiv  2 (\sigma_{\mu \nu}  - \omega_{\mu \nu}   )  \, . ,
\eea  
where $R_{\mu \nu}$ is the Ricci curvature tensor, and $C_{\rho \mu \sigma \nu}$ is the Weyl curvature tensor. The operator $D_\mu$ is the spatial covariant derivative, which is defined through its action on an arbitrary tensor field:   $D_{\mu} T_{\nu_1 ,  .. , \nu_n }^{\qquad  \gamma_1 , .. , \gamma_m } \equiv  h_{ \nu_1 }^{\, \alpha_1 } .. h_{ \nu_n }^{\, \alpha_n }    \,  h_{ \beta_1 }^{\, \gamma_1 } .. h_{ \beta_m }^{\, \gamma_m }    \, h_{ \mu }^{\, \sigma } \nabla_\sigma  T_{\alpha_1 ,  .. , \alpha_n }^{\qquad \beta_1 , .. , \beta_m }$ .

Regarding the decomposition in \eqref{Pimultikappa}, we note that  the truncation of the multipole series at second order in the direction variables $\bm{e}$ and $\bm{\kappa}$ of the photon congruence is \emph{exact} for any spacetime description. 
The coefficients of the series are constructed from the kinematic variables of the observer congruence together with the Ricci focusing term $u^\mu u^\nu R_{\mu \nu}$ and the electric part of the Weyl tensor $u^\rho u^\sigma  C_{\rho \mu \sigma \nu}$. 

\subsection{Lema\^{\i}tre-Tolman-Bondi spacetime}
\label{sec:LTB} 
We now consider the special case of the spherically-symmetric LTB spacetime metric \cite{lemaitre,tolman,bondi} (see e.g. the  books \cite{b_k_h_c_2009,plebanski_krasinski_2006} for an introduction). 
We write the LTB line element in spherical coordinates $x^\mu = (t,r,\theta,\phi)$ adapted to the center of the LTB structure as 
\bea
\label{metricLTB}
\hspace*{-0.35cm}  \d s^2 = - \d t^2 + R(t,r)\d r^2 + A^2(t,r)\! \left(\d \theta^2 + \sin^2(
\theta)\d \phi^2 \right)  , 
\eea  
with $R(t,r) \equiv (\partial_r A(t,r) )^2/(1-k(r))$, where $k(r)$ specifies the spatial Ricci curvature of the LTB model \cite{Marra:2011ct} and reduces to a constant times $r^2$ in the FLRW spacetime limit. 
The metric is required to be a solution to the Einstein equation $R_{\mu \nu} - R g_{\mu \nu}/2 = 8\pi G T_{\mu \nu}$ with a dust source, such that the energy momentum tensor reads $T_{\mu \nu} = \rho \delta^{t}_{\mu} \delta^{t}_{\nu}$; for the explicit form of the independent components of Einstein's equations, see section~\ref{sec:paramLTB} where we also specify the LTB solution used in our analysis. 
We consider an observer congruence that is comoving with the foliation of the metric representation in \eqref{metricLTB}, such that $u^\mu = \delta^\mu_t$. 
It follows immediately that $\omega_{\mu \nu} = 0$ and $a^\mu = 0$ in the kinematic decomposition \eqref{def:expu}, 
and the multipole coefficients in \eqref{Picoefkappa} read 
\bea
\label{PicoefLTB}
&& \Sigma^{\it{o}} = - \frac{1}{3} u^\mu u^\nu R_{\mu \nu}         \, , \qquad   \Sigma^{\bm{e}}_\mu  =   0 \, ,  \nonumber  \\   
&& \Sigma^{\bm{ee}}_{\mu \nu} =       -  u^\rho u^\sigma  C_{\rho \mu \sigma \nu}     \,   , \qquad  \Sigma^{\bm{e\kappa}}_{\mu \nu} =  2 \sigma_{\mu \nu}   \quad \text{(LTB)}  \, , 
\eea  
which can be straightforwardly computed in terms of the LTB metric components \eqref{metricLTB} and their gradients. We list the multipole terms for the LTB metric in appendix \ref{sec:LTBmultipole} for convenience.  

In the case of a radially propagating congruence of photons, $\bm{\kappa}$ vanishes, and the redshift drift signal is determined solely from Ricci focusing and electric Weyl curvature. 
In general, however, the propagation of photons with a non-zero impact parameter relative to the LTB structure will give rise to the additional terms $- \kappa^\mu \kappa_\mu$ and $e^\mu   \kappa^\nu \Sigma^{\bm{e\kappa}}_{\mu \nu}  = 2 e^\mu   \kappa^\nu  \sigma_{\mu \nu}$ in \eqref{Pimultikappa}. 
In the FLRW limit, the only non-zero coefficient is the Ricci focusing term and the integrand \eqref{Pimultikappa} reduces to $\Pi \! \overset{\mathrm{FLRW}}{\rightarrow}\!\!\!  - \frac{1}{3} u^\mu u^\nu R_{\mu \nu}$.  

\section{Model setup and light propagation} 
\label{sec:num} 
In this section we describe the details of our numerical analysis. 
In section~\ref{sec:paramLTB} we specify the LTB model that we investigate. 
In section~\ref{sec:init} we detail the geodesic equations for light propagation and discuss the initial conditions used for specifying the light beams. 

\subsection{Parameterization of the Lema\^{\i}tre-Tolman-Bondi structure} 
\label{sec:paramLTB} 
The solution of the LTB metric specified in section~\ref{sec:LTB} is determined by two independent components of the Einstein field equation which can be integrated to yield  
\bea
\label{EE1}
(\partial_t A(t,r))^2 = \frac{2M(r)}{A(t,r)} - k(r) \, , 
\eea 
and 
\bea
\label{EE2}
\frac{\partial_r M(r)}{4\pi G A^2(t,r) \partial_r A(t,r)} = \rho \, , 
\eea 
where the integration constant $M(r)$ is the active gravitational mass inside a shell of radius $r$ of the LTB structure. 
Equation \eqref{EE1} can be solved for $A(t,r)$ for valid specifications of the functions $M(r)$ and $k(r)$, provided initial conditions for $A(t,r)$. 
We impose that the big bang happens synchronously in the model by requiring that the big bang function 
\bea 
\label{bang}
t_B(r) = t \, - \!\! \int\limits_0^{A(t,r)} \!\!\! \mathrm{d}\tilde{A} \,  \frac{1}{\sqrt{\frac{2M(r)}{\tilde{A}} - k(r) }} 
\eea 
is zero for all $r$\footnote{Note that this integral can be solved explicitly. Solutions are given in e.g. \cite{b_k_h_c_2009} but for our work we found it convenient to solve the ODE \eqref{EE1}.}. We furthermore choose the spatial curvature profile such that 
\bea 
\label{kdef}
    k(r) = 
\begin{cases}
    -1.3 \times 10^{-7} r^2\left( \left( \frac{r}{r_b} \right)^m - 1 \right)^6 ,& \text{if } r < r_b\\
    0 \; ,              & \text{otherwise} \, , 
\end{cases}
\eea 
where $r_b$ is the radius of the LTB structure, outside of which the curvature is that of an Einstein de-Sitter (EdS), which we shall refer to as the background metric.
The condition $t_B(r) = 0$ and the profile \eqref{kdef} yields a closed-form solution to \eqref{EE1} in terms of $t,r$ and $M(r)$; see ~\cite{VanAcoleyen:2008cy} or Appendix~A in~\cite{Redlich:2014gga}. 
The function $M(r)$ can be specified through a suitable choice of initial conditions for $A(t,r)$ (~\cite{VanAcoleyen:2008cy}); which in turn specifies $A(t,r)$ throughout. 
Here we choose the EdS-adapted initial conditions with  $A(t_i,r)=a_{\text{EdS}}(t_i) r$, where $a_{\text{EdS}}(t_i) = (t_i/t_{0})^{2/3}$ and $t_{0} = 2/3/H_{0}$, with initial scalefactor  $a_{\text{EdS}}(t_i) = 1/1100$ and with the initial time $t_i$ fixed by $H_{0} = 70$km/s/Mpc. 

The curvature model \eqref{kdef} represents a central void surrounded by a steep overdensity. We used $r_b = 40$Mpc for all numerical computations. The choice $m = 6$ was used during the main part of our study, but to test the significance of the exact density profile on our results, we have also studied a single light ray passing through a structure with $m = 2$. In addition, we have made minor tests using different void depths  and sizes of the surrounding overdensity by scaling the function $k(r)$.  We find that the results do not qualitatively depend on the exact choice of density profile but note that for a more significant change in density profile, the results should be expected to be even qualitatively different. This is for instance seen by the comparisons in the appendix of ~\cite{Koksbang:2015ctu}, which reveal that a prominent deviation from the FLRW result can be expected if the LTB inhomogeneity does not reduce {\em exactly} to an FLRW background at a reasonably small value of $r$. We also note that the strongest signs of inhomogeneity appear at the edges of the LTB structure, where the density contrast is at its steepest.
\newline\indent

\subsection{Light propagation and initial conditions} 
\label{sec:init} 
We choose a comoving observer located in the EdS region of the spacetime with worldline passing through the point $\obs$ given by time coordinate $t_\obs \! =\! 2/3/H_{0}$ with $H_{0} \! =\! 70$~km/s/Mpc, corresponding to the present time in our model. The radial coordinate is chosen such that the observer is located $10$~Mpc outside of the structure: $r_\obs = r_b + 10$~Mpc, and the angular coordinates $\theta_\obs, \phi_\obs$ are fixed arbitrarily.  

For each central null ray\footnote{We use the term \sayy{central null ray} to describe the light ray which passes between the primary points of emission $\emi$ and observation $\obs$, and around which we shall consider the extension into a congruence of null rays (see below).}, we choose a random line of sight with equal probability for all directions on the observer's sky. 
Upon transformation to spherical coordinates, this determines initial values of $k^r, k^{\theta}$ and $k^{\phi}$ (and we always require $k^r>0$).  With this procedure, some light rays will miss the LTB structure and only propagate through EdS spacetime. We remove these rays from the analysis so that they do not e.g. contribute to computations of mean quantities. The spatial direction vector is normalized in accordance with the initial condition of $k^t$ which can be chosen arbitrarily without loss of generality. 
To compute the components of $\bm{\kappa}$ along the central null ray, we need the partial derivatives of the tangent vector along the ray. We use the procedure detailed in \cite{Koksbang:2020zej}, i.e. we solve
\bea 
\label{kprime}
\frac{dk^{\mu}_{,\nu}}{d\lambda} = \frac{\partial}{\partial x^{\nu}}\frac{dk^{\mu}}{d\lambda} - k^{\beta}_{,\nu}k^{\mu}_{,\beta} 
\eea 
simultaneously with the geodesic equation
\bea 
\label{geodesic}
\frac{d}{d\lambda}\left( g_{\alpha\beta}k^{\beta}\right) = -\frac{1}{2}g_{\mu\nu,\alpha}k^{\mu}k^{\nu}.
\eea 
The solution to the system of equations \eqref{kprime}, \eqref{geodesic} is specified by the initial conditions for $k^\mu$ described above, along with initial conditions for $k^\mu_{,\nu}$. These initial conditions uniquely define a 4-dimensional congruence of null geodesic rays around the central null ray.

We are interested in computing the redshift drift corresponding to comoving emitters, but we do not a priori know the initial conditions for $k^\mu_{,\nu}$ that correspond to a comoving emitter passing through a given event of emission $\emi$ along the central null ray.\footnote{In principle we could solve for the appropriate photon congruence description connecting a given emitter worldline with the observer worldline, by solving the geodesic deviation equation with Dirichlet boundary conditions as specified in section~3.1 of~\cite{Korzynski:2017nas}. However, in practice it is computationally heavy to solve this boundary value problem for each point along the central null ray. } 
We thus follow an empirical approach, where we first define a bundle of null rays, and then assess whether emitters of the incoming light rays on the observer worldline (almost) correspond to comoving emitters. 
We shall for this purpose choose initial conditions for the null bundle such that $\bm{\kappa}_\obs = \bm{0}$, i.e., the emitting sources are constrained to remain at a fixed direction on the observer's sky.
This choice of initial condition is compatible with setting  $k^{\mu}_{,t}|_\obs = \frac{1}{k^t}\frac{dk^\mu}{d\lambda}|_\obs$ in the LTB adapted coordinate system. The remaining initial conditions $k^{\mu}_{,i}|_\obs$ must be compatible with this choice and the geodesic requirement \eqref{geodesic}, but are otherwise gauge choices of the signal arriving at the observer worldline.\footnote{See appendix~\ref{sec:GD} for further discussions on the gauge choices involved with the initialisation of the geodesic null bundle.} 
Following~\cite{Koksbang:2020zej}, we set $k^\mu_{,i}|_\obs = 0$ in Cartesian coordinates before making a coordinate transformation to spherical coordinates. We summarise our choice of initial conditions as follows  
\bea 
\label{initialkprime}
k^{\mu}_{,t}|_\obs =  \frac{1}{k^t}\frac{dk^\mu}{d\lambda}\Big|_\obs \, , \qquad k^\mu_{,i}|_\obs = 0 \, , \qquad i = x, y, z.
\eea  
When the light rays travel exclusively in the FLRW region, the intitial conditions \eqref{initialkprime} are compatible with comoving sources as emitters of the signal. 
However, once light enters the LTB structure, these initial conditions will generally not be compatible with comoving emitters, since only radial light rays are repeatable in LTB models (see e.g. ~\cite{Krasinski:2010rc}). 
However, emitters of the light contained in the bundle might nevertheless be \emph{close} to being comoving.


As discussed in detail in appendix~\ref{sec:GD}, we can determine the family of 4-velocity fields of sources which are candidates for having emitted the light with incoming conditions \eqref{initialkprime} at the observer.  
The 4-velocity of the source can be chosen uniquely from specifying $\alpha$ in \eqref{devdec}. Here we make the following choice of 4-velocity  
\bea
\label{nchoice}
n^\mu \equiv \frac{X_{\text{scr}}^\mu}{\sqrt{ - g_{\nu \rho} X_{\text{scr}}^\nu X_{\text{scr}}^\rho}} \, , \quad  X_{\text{scr}}^\mu = \tilde{X}^\mu  + \alpha_{\text{scr}} k^\mu \, , 
\eea 
where $\bm{\tilde{X}}$ is determined by \eqref{Liedeviation2}, and  
where $\alpha_{\text{scr}} \equiv \tilde{X}^\mu e_\mu / E_u$. We label the photon energy $E_u \equiv - k^\mu u_\mu$ with a subscript from now on, to distinguish the energy measured in the LTB comoving frame from the energy as measured in other frames. 
This choice of sources ensures that the spatial direction of propagation of the photon $\bm{e}$ as seen in the comoving LTB frame is indeed also spatial in the frame of the source, i.e., $n^\mu e_\mu = 0$. Thus, any difference between $\bm{u}$ and $\bm{n}$ is due to the components of $\bm{u}$ in the screen space orthogonal to the two dimensional congruence of light spanned by $\bm{k}$ and $\bm{\tilde{X}}$; hence the use of the subscript $\text{scr}$ which is short for \sayy{screen space}.

\begin{figure}[h]
	\centering
	\includegraphics[scale = 0.5]{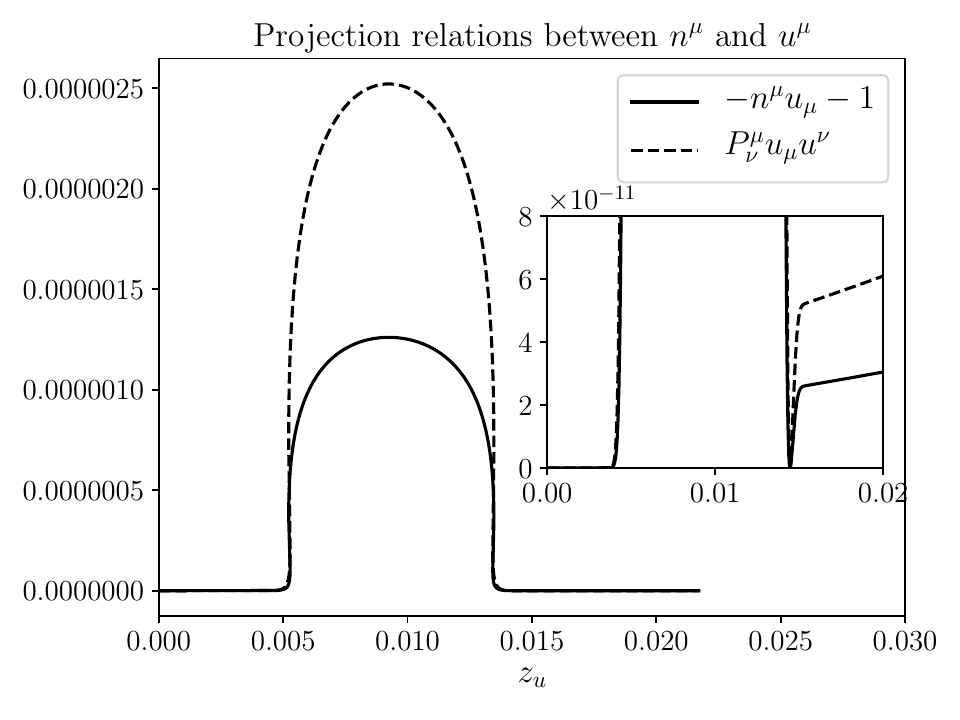}
	\caption{Projection of $u^\mu$ orthogonal to $n^\mu$ and $k^\mu$ using the projection tensor together with projection of $u^\mu$ along $n^\mu$. along a single light ray. }
	\label{fig:projection}
\end{figure}

\begin{figure}
	\centering
	\includegraphics[scale = 0.5]{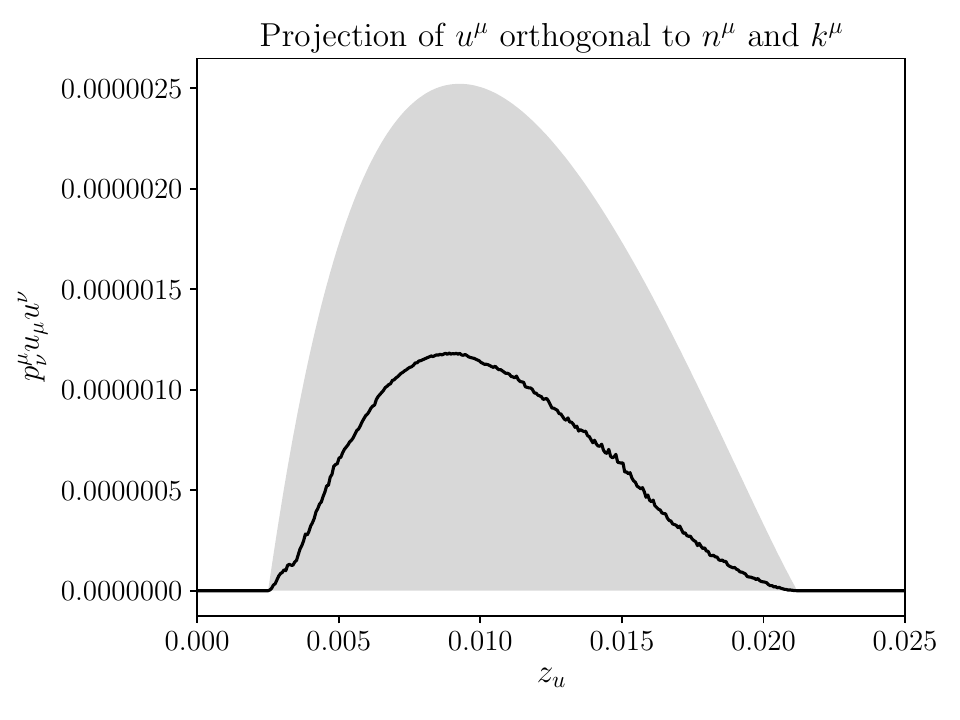}
	\caption{Projection of $u^\mu$ orthogonal to $n^\mu$ and $k^\mu$ using the projection tensor. The black line indicates the mean over 1400 light rays while the grey-shaded area indicates the spread. The result is plotted against the redshift of comoving emitters. }
	\label{fig:P}
\end{figure}

 The emitter 4-velocity field $n^\nu$ turns out to be very close to the comoving 4-velocity field $u^\mu$ at every point along the central null ray. 
 This can be seen by computing the norm of $\bm{u}$ in the screen space orthogonal to $\bm{n}$ and $\bm{k}$: $P^{\mu \nu} u_\mu u_\nu$, with $P^\mu_{\, \nu} \equiv\frac{- k^\mu k_\nu}{E_n^2} + \frac{k^\mu n_\nu}{E_n} + \frac{n^\mu k_\nu}{E_n} + g^\mu_{\, \nu}$. Alternatively, we could compute the relative tilt $-n^\mu u_\mu$. We show both closeness-measures in figure \ref{fig:projection} for a fiducial light ray. As seen, the two measures are very similar and indeed in general differ by a factor of two at lowest order\footnote{It can be verified that $P^{\mu \nu} u_\mu u_\nu = v^\mu v_\mu + \mathcal{O}(v^3)$, where $v^\mu$ is the relative velocity defined through $u^\mu = (n^\mu + v^\mu)/\sqrt{1 - v^\nu v_\nu}$. Thus, we have the following relation $P^{\mu \nu} u_\mu u_\nu = 2(- n^\mu u_\mu - 1) + \mathcal{O}(v^3)$} in $\bm{v}$; therefore, we shall analyse the former measure only for the full set of null rays.   
 The projection $P^{\mu \nu} u_\mu u_\nu$ is shown for 1400 light rays in figure~\ref{fig:P}.  
 As seen, the projection is small -- at most of order $10^{-6}$, and of order $10^{-10}$ for emitters in the FLRW region (on the opposite side of the structure as compared to the observer) -- which means that the relative velocity between $\bm{n}$ and $\bm{u}$ is at most of order  $10^{-3}$, and reaches levels of order $10^{-5}$ once the structure has been traversed. Thus, comoving emitters are {\em close to} being emitters of the light signal received at the observer worldline, even when being situated within the structure. 
We thus expect the redshift drift signal in the comoving frame to be close to that in the frame of $\bm{n}$. 
\newline\newline  
In the following we shall analyse the redshift drift signal in the frame of the almost-comoving geodesic observers generated by the 4-velocity $\bm{n}$.
Similarly to the expression for the redshift drift used in ~\cite{Koksbang:2020zej} we can compute the redshift drift corresponding to the geodesic emitter with 4-velocity $n^{\mu}$   
as (see appendix~\ref{sec:GD} for details)
\bea
\label{zdriftn}
\delta z_n \equiv -\delta t_\obs \cdot \left( \frac{E_n  }{E_\obs^2}k^t_{,t} |_\obs + \frac{1}{E_n}n^\mu n^\mu \nabla_\mu k_{\nu}   \right),
\eea
where $E_n\equiv -k^\mu n_\mu$, and where evaluation is at any point along the central null ray. 
 
We shall in addition make use of the following convenient approximation of $\delta z_n$: 
\begin{align}\label{zdriftu}
\begin{split}
\delta z_u &\equiv -\delta t_\obs \cdot \left( \frac{E_u  }{E_\obs^2}k^t_{,t} |_\obs + \frac{1}{E_u}u^\mu u^\mu \nabla_\mu k_{\nu}   \right)\\
 &= \frac{\delta t_\obs}{k^t_\obs} \left( -(1+z_u) k^t_{,t|\obs} + \frac{k^t_{,t|\emi}}{1+z_u} \right),
\end{split}
\end{align}
which we, in the following section, shall verify remains close to $\delta z_n$. 
The approximation \eqref{zdriftu} can conveniently be written as the integral representation \eqref{redshiftdriftint} with integrand \eqref{Pimultikappa} and coefficients \eqref{PicoefLTB}.  

\section{Results}\label{sec:results}
In this section we present the results obtained by propagating light rays through a single LTB structure, as described in section~\ref{sec:num}.
In section~\ref{sec:singleray} we present results obtained by considering a single random light ray. Afterwards we move on to present results obtained from 1400 light rays in section~\ref{sec:multiplerays}. 

\subsection{Single light ray} 
\label{sec:singleray}

\begin{figure}
\centering
\includegraphics[scale = 0.5]{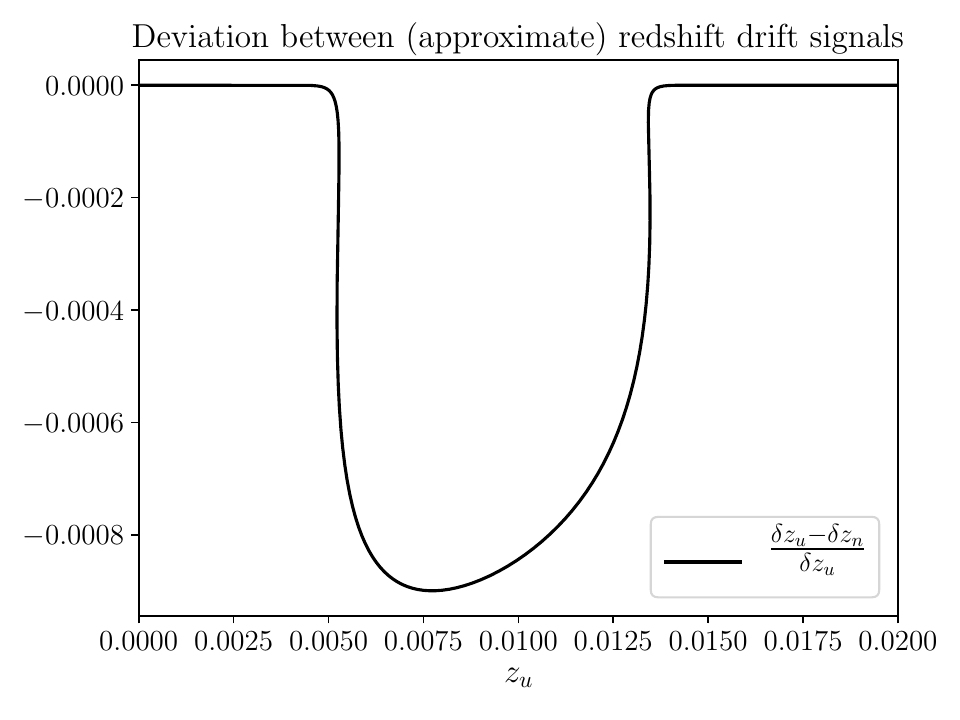}
\caption{Deviation between two redshift drifts along a fiducial light ray.}
\label{fig:Deltadelta}
\end{figure}

In this section, we show results from propagating a single light ray with a random impact parameter through the LTB structure. 
We set $\delta t_\obs = 30$ years, where $\delta t_\obs$ is the observer's proper time elapsed between two measurements of the redshift and $\delta z$ is the drift (change) in the redshift of a source during that interval. 
We show results using the density profile defined by $m = 6$ but have verified that the results are similar for the profile corresponding to $m = 2$ as well as for models with different scalings of $k(r)$ to enhance/suppress the structure.
\newline\newline 
We compute the redshift drift signal in the frame of the geodesic and almost-comoving emitters with 4-velocity field $\bm{n}$, as detailed in section~\ref{sec:init}, and compare the exact redshift drift signal of these emitters \eqref{zdriftn} to the approximation  \eqref{zdriftu}.
This comparison is shown in figure \ref{fig:Deltadelta} where it is seen that the deviation between $\delta z_u$ and $\delta z_n$ is maximally of order $10^{-3}$, as is also expected based on figure \ref{fig:P}.
We thus find that $\delta z_u$ is a good approximation of $\delta z_n$ along the entire null ray -- a result which we verify to hold for the full sample of null rays considered in our analysis -- and we use $\delta z_u$ as a convenient approximation of the redshift drift signal in the following.    

\begin{figure}
\centering
\includegraphics[scale = 0.5]{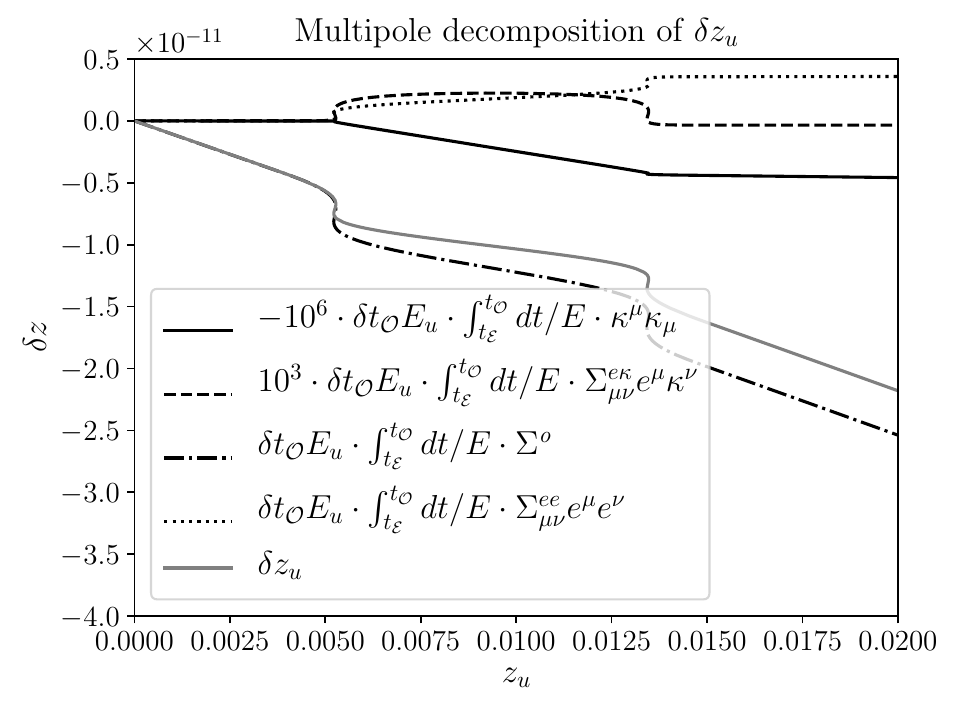}
\caption{Individual components of the redshift drift signal along a light ray. The two subdominant components are scaled to ease assessment of their qualities.}
\label{fig:singe_components}
\end{figure}

\begin{figure}
\centering
\includegraphics[scale = 0.5]{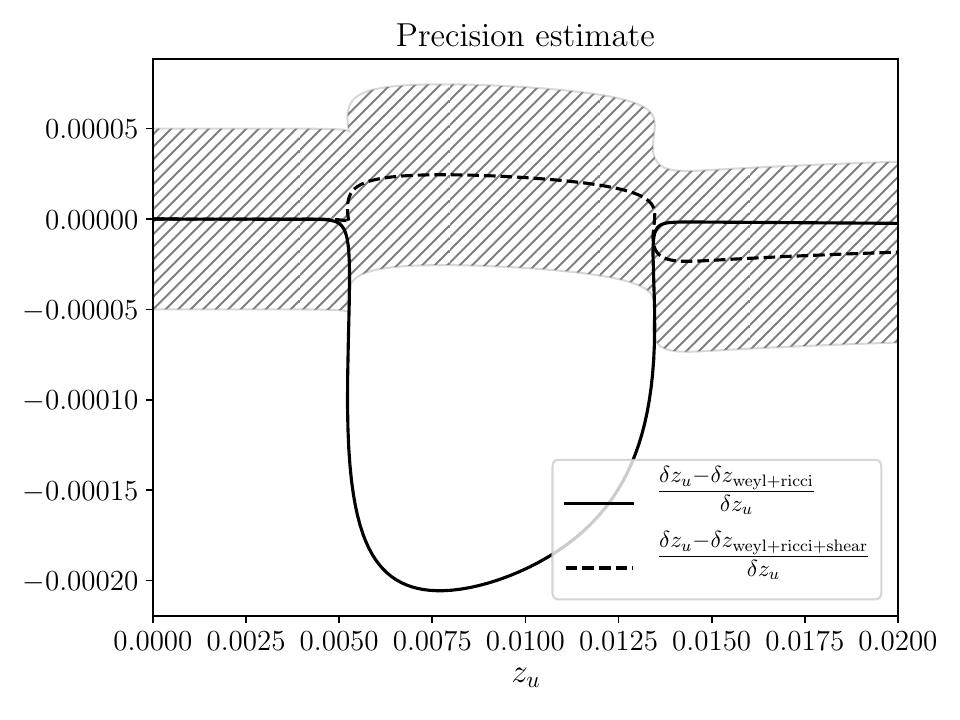}
\caption{Deviations between $\delta z_u$ and two approximations computed by including the two and three most dominant 
multipole contributions -- either Weyl+Ricci or Weyl+Ricci+shear components, corresponding to  $\delta z_{\rm Weyl+Ricci}\equiv E\int_{t_\emi}^{t_\obs}dt/E\Sigma^{\it{o}}+\Sigma^{\bm{ee}}_{\mu \nu}e^{\mu}e^{\nu}$ and $\delta z_{\rm Weyl+Ricci+shear}\equiv E\int_{t_\emi}^{t_\obs}dt/E\Sigma^{\it{o}}+\Sigma^{\bm{ee}}_{\mu \nu}e^{\mu}e^{\nu} + \Sigma^{\bm{e\kappa}}_{\mu \nu} =  2 \sigma_{\mu \nu}e^\mu \kappa ^\nu $, respectively. The shaded area indicates a rough estimate of the numerical precision of the computations, corresponding to 4 significant digits.}
\label{fig:precision}
\end{figure}

Figure \ref{fig:singe_components} shows the contributions of the multipole components in the representation \eqref{Pimultikappa} together with the total redshift drift signal approximation \eqref{zdriftu},  
with initial conditions for the light bundle as specified in section~\ref{sec:init}. 
We immediately see that the two main contributions are those corresponding to the Ricci and Weyl tensors. This is encouraging since these two terms do not depend on the extension of the congruence 
$k^\mu_{,\nu}$ away from the central null ray, and are thus much more easily computed than the (integral of the) two terms $- \kappa^\mu \kappa_\mu$ and $e^\mu   \kappa^\nu \Sigma^{\bm{e\kappa}}_{\mu \nu}  = 2 e^\mu   \kappa^\nu  \sigma_{\mu \nu}$ in \eqref{Pimultikappa}.
\newline\indent 
The contribution from the drift $\kappa$ is actually so small that it is  nearly swamped by the numerical errors of the computations 
which make our redshift drift determinations reliable only at the level of 4 significant digits. 
With this precision, we are just barely able to see the contribution from the next-smallest term, the shear term. This is illustrated in figure~\ref{fig:precision} where we show the relative difference 
between $\delta z_u$ and the approximate redshift drift computed using only the two or three dominant multipole contributions, respectively. 
We see that the approximation $\delta z_{\rm Weyl+Ricci+shear}\equiv E\int_{t_\emi}^{t_\obs}dt/E\Sigma^{\it{o}}+\Sigma^{\bm{ee}}_{\mu \nu}e^{\mu}e^{\nu} + \Sigma^{\bm{e\kappa}}_{\mu \nu} =  2 \sigma_{\mu \nu}e^\mu \kappa ^\nu$, using the three dominant multipole contributions, gives an accurate determination of $\delta z_u$ within the numerical errors, a rough estimate of which are shown as a shaded area.
The integral-term involving $- \kappa^\mu \kappa_\mu$ is roughly two orders of magnitude smaller than what can be resolved within the numerical errors of this analysis, and might thus be neglected for all practical purposes.

\begin{figure*}
\centering
\subfigure{
\includegraphics[scale = 0.5]{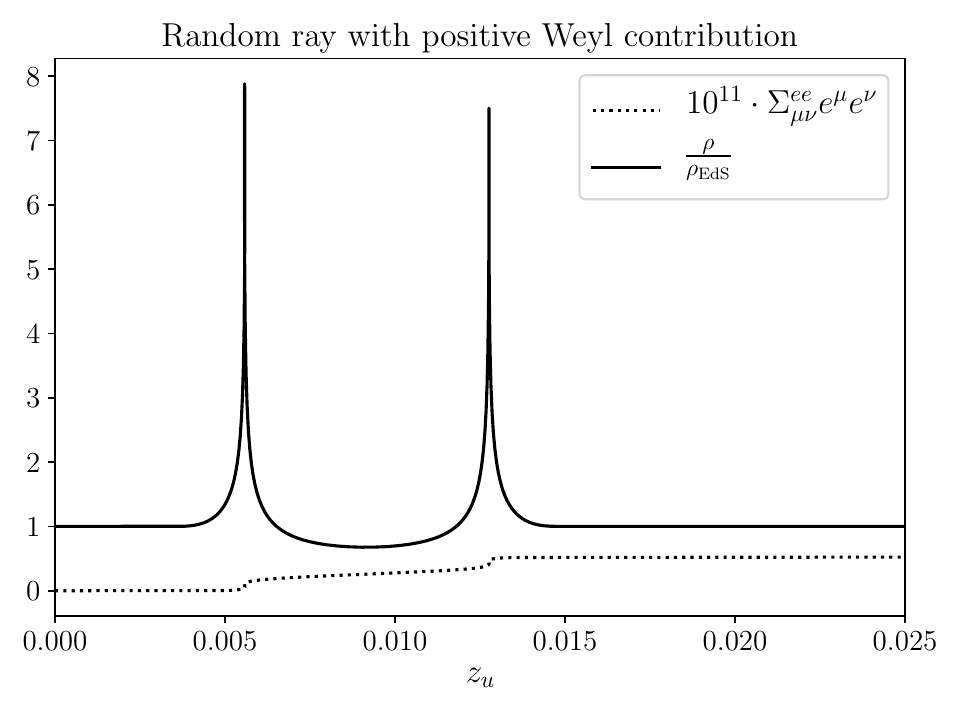}
}
\subfigure{
\includegraphics[scale = 0.5]{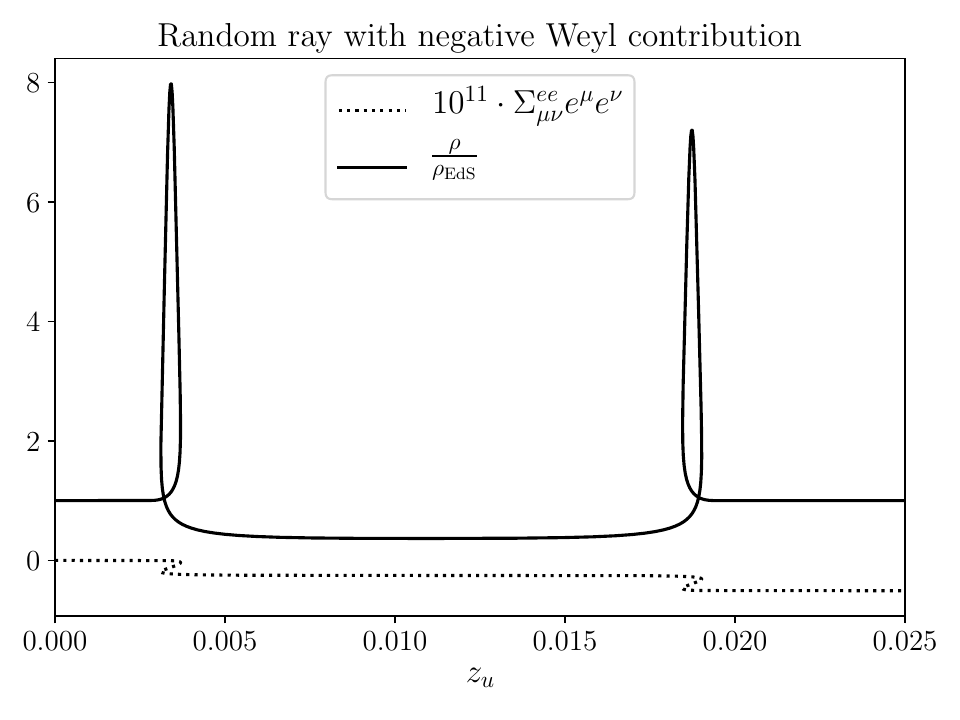}
}
\caption{Weyl contributions to the redshift drift along fiducial light rays with opposite sign of the Weyl contribution. The contributions are shown together with the density profiles along the light rays.}
\label{fig:PositiveNegative}
\end{figure*}

We note that the Weyl contribution to the redshift drift signal is significant -- also for emitters placed outside of the LTB structure (on the opposite side of the observer).  
However, the Weyl contribution can be both positive and negative, 
so the mean Weyl contribution may be modest when averaging over many individual light rays with different impact parameters. 
Figure \ref{fig:PositiveNegative} shows the Weyl contribution along two arbitrarily chosen fiducial rays 
with positive and negative Weyl contributions, respectively. The contributions are shown together with the density profile along the individual rays to illustrate that the negative Weyl contribution appears to occur when light rays move further into the underdense region of the structure. Although it is not clearly visible in the figure, we note that the redshift is non-monotonous along both rays.

\begin{figure}
\centering
\includegraphics[scale = 0.5]{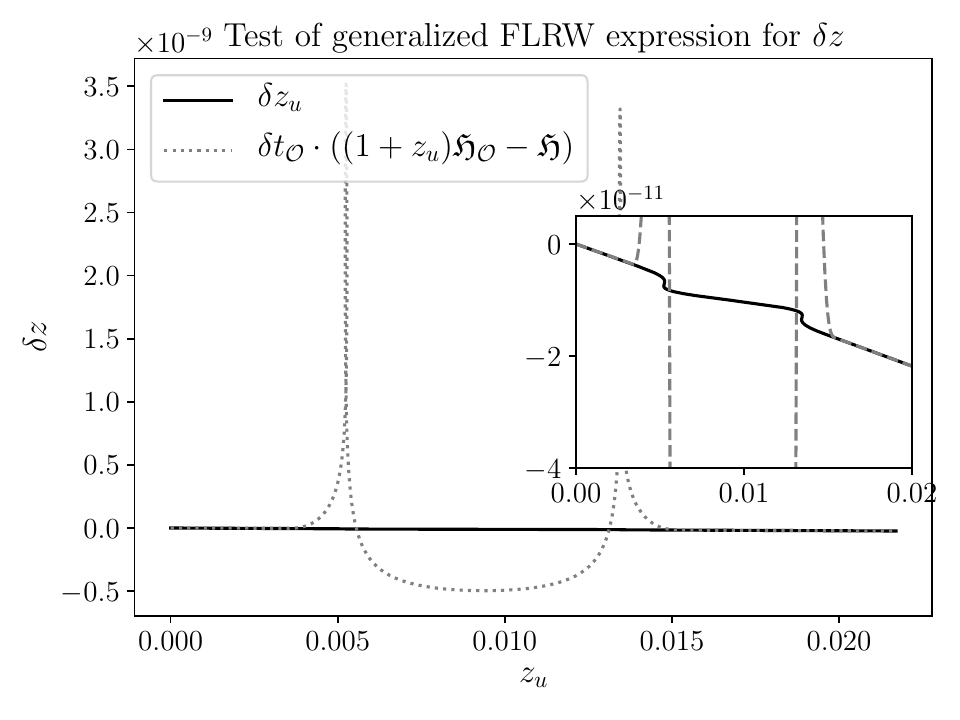}
\caption{Actual redshift drift compared with simple expectation based on generalized FLRW relation along a single light ray. A close-up is included since the simple approximation is several orders of magnitude larger than the actual redshift drift several places along the light rays.}
\label{fig:generalizedFLRW}
\end{figure}

Before moving on to discuss the results obtained when averaging over several light rays, we note that a simple relation between the redshift drift and the local expansion rate along light rays is {\em not} apparent. 
Indeed, in the absence of certain systematic impacts of anisotropies along the central null ray, the redshift drift signal is expected to simplify to an expression similar to the FLRW relation~\cite{Heinesen:2020pms}
\begin{align} 
\label{eq:dzsimple}
    \delta z_{\rm simple} \equiv \delta t_\obs \left( (1+z) \Eu_\obs - \Eu \right) ,
\end{align}
with the generalized \sayy{Hubble parameter} $\Eu \equiv\frac{1}{3}\theta + e^\mu e^\nu \sigma_{\mu\nu}$ describing the rate of expansion of length scales along the direction of the photon 4-momentum.  
However, figure \ref{fig:generalizedFLRW} shows that \eqref{eq:dzsimple} is not a good approximation for emitters located within the LTB structure.  The approximation $\delta z_{\rm simple}$ departs from $\delta z_u$ by orders of magnitude for most emitters located within the LTB structure, which might be assigned to both the large departures of $\theta$ from the EdS background expansion rate for most points within the LTB structure and to the projected shear contribution within the structure. This is illustrated in figure \ref{fig:singe_components} along a fiducial light ray. For emitters located in the FLRW region (on the opposite site of the structure from the observer), \eqref{eq:dzsimple} provides an extremely good approximation as it is simply the background EdS redshift drift which $\delta z_u$ reduces to outside of the structure, to the precision of our computations (around 5 significant digits).

\subsection{Multiple light rays} 
\label{sec:multiplerays}
\begin{figure}
\centering
\includegraphics[scale = 0.5]{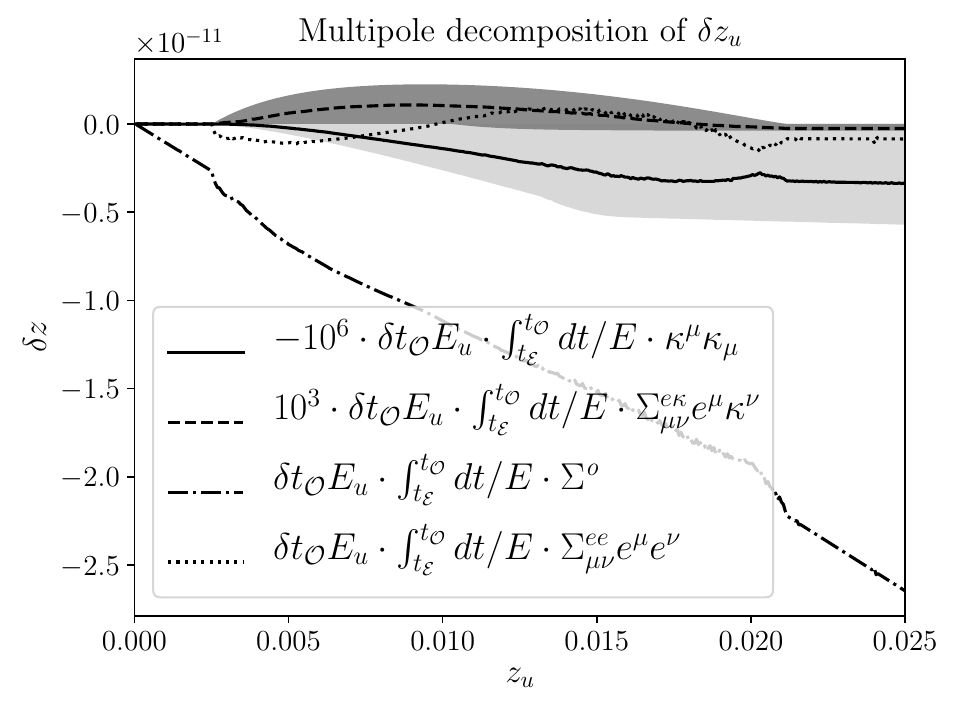}
\caption{Mean of individual components of the redshift drift signal along 1400 light rays. The two subdominant components are scaled to ease assessment of their qualities and are shown together with their spreads (shaded areas).}
\label{fig:components_1500}
\end{figure}

\begin{figure}
\centering
\includegraphics[scale = 0.5]{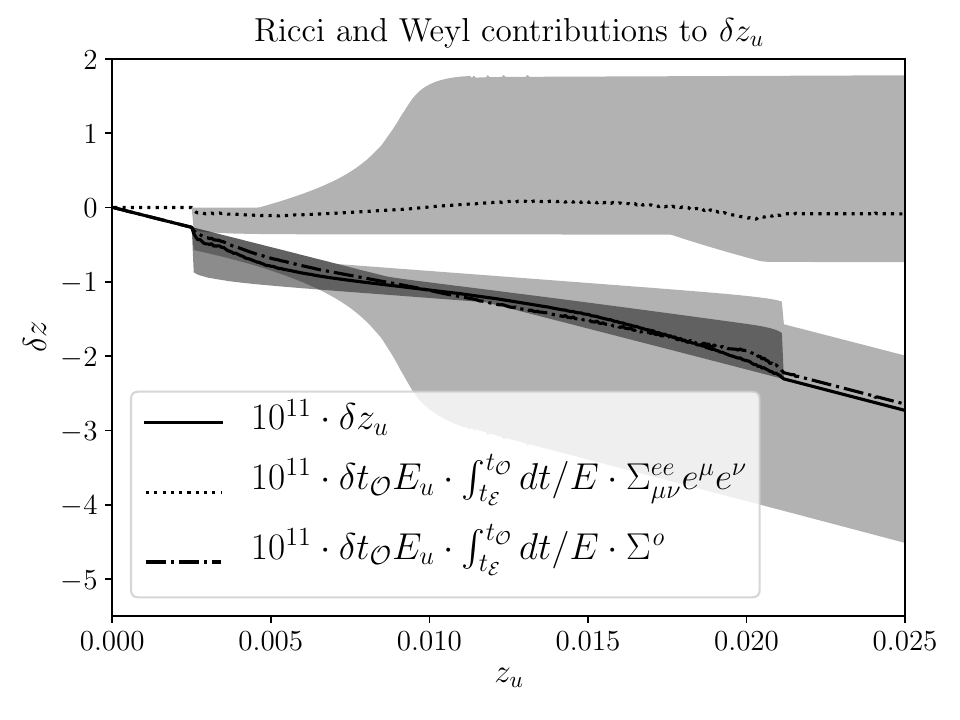}
\caption{Ricci and Weyl contributions to the redshift drift along 1400 light rays. The black lines indicate mean values while the shaded areas indicate the spread.}
\label{fig:ricciweyl_1500}
\end{figure}
In this section, we redo the analysis for 1400 light rays and compute the mean value and spread of the results. The observer is always the same (placed at $t = t_\obs$ and $r = r_b + 10$Mpc). Each light ray is propagated until it reaches $z = 0.025$, which is enough to traverse the entire LTB structure.
\newline\newline
Figure~\ref{fig:components_1500} shows the integrated multipole components of the redshift drift. It is visible from the figure that the two components which depend on the drift of the viewing angle, $\bm{\kappa}$, 
are sub-dominant and can to a high precision be neglected, as we also found for the single light ray above. 
This means that we can to a good precision approximate the redshift drift signal from the Ricci and electric Weyl curvature components along the individual null rays.
The figure also shows that the Weyl contribution does not vanish on average after traversal of the light ray through the structure, but the remaining mean effect is much smaller than the contribution from the average Ricci term, with the mean of the former making up approximately 5 \% of the signal after traversal of the entire structure. This is also seen in figure \ref{fig:ricciweyl_1500}, which shows mean and spread of the Ricci and Weyl contributions together with the total redshift drift signal.
\begin{figure}
\centering
\includegraphics[scale = 0.5]{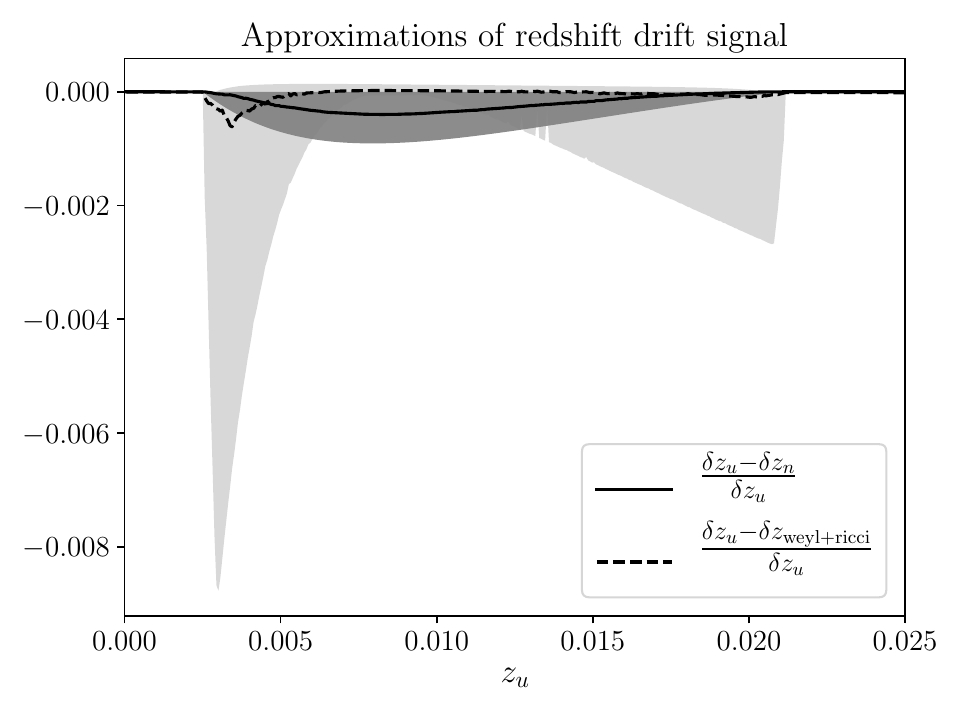}
\caption{Deviation between redshift drifts $\delta z_u$ and $\delta z_n$ as well as between $\delta z_u$ and the redshift drift computed without the $\bm{\kappa}$-contributions, i.e. $\delta z_{\rm Weyl+Ricci}\equiv E\int_{t_\emi}^{t_\obs}dt/E\Sigma^{\it{o}}+\Sigma^{\bm{ee}}_{\mu \nu}e^{\mu}e^{\nu}$. The black lines indicate mean values while the spreads are indicated by shaded areas.}
\label{fig:diff}
\end{figure}
\begin{figure}
\centering
\includegraphics[scale = 0.5]{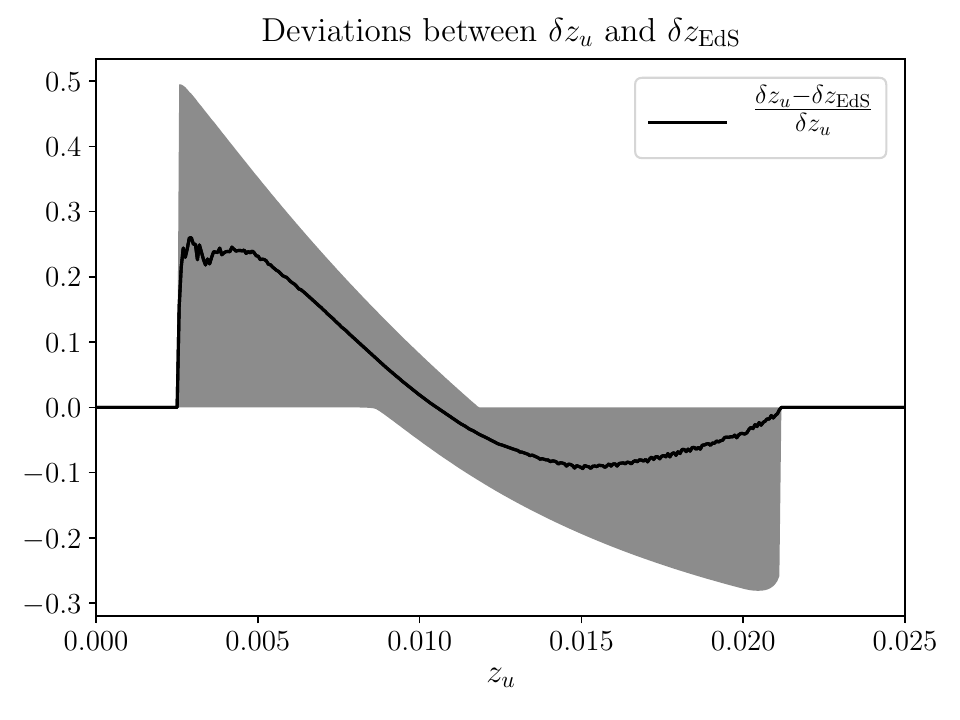}
\caption{Deviation between $\delta z_u$ and the background (EdS) redshift drift. The line indicates the difference between the EdS redshift drift and the mean redshift drift while the shaded area indicates the spread. }
\label{fig:diff_bg}
\end{figure}
\newline\indent
In figure \ref{fig:diff} we show the deviations between $\delta z_u$ and the signal corresponding to the redshift drift without the 
$\bm{\kappa}$-contributions as well as the difference between $\delta z_u$ and $\delta z_n$. Both of the differences are subpercentage, again indicating that we can to a good approximation treat the sum of the Ricci and Weyl contributions to the redshift drift as {\em the} redshift drift signal measured by a comoving observer in the FLRW region, and emitted by comoving sources along the light paths. 
\newline\indent
Lastly, figure \ref{fig:diff_bg} shows the difference between the mean redshift drift and the EdS (background) redshift drift. The difference becomes quite large for typical emitters of light, 
emphasizing the potential importance of taking effects of structures into account when interpreting real upcoming redshift drift data. 
However, we note that the relative departures from the EdS signal are expected to decrease for longer distances of light propagation than a single LTB structure.

For the comparison in figure \ref{fig:diff_bg} we compute the EdS redshift drift as
\begin{align}
    \delta z_{\rm EdS} = \left(1+z_u (\lambda_\emi) \right)H_0 - H(z_u(\lambda_\emi)) \, ,  
\end{align} 
where $H(z_u)$ is the background EdS Hubble parameter function as parameterised in terms of EdS redshift, $z_{\rm EdS}$, and evaluated at the value $z_{\rm EdS} = z_u$. 
We note that, since the local difference between $z_{\rm EdS}(t(\lambda))$ and $z_u(\lambda)$ is subpercentage along the null rays, the result does not change significantly if we instead use an EdS parameterization in terms of the time-parameter of emission:
\begin{align}
    \delta z_{\rm EdS}^\text{alternative}  = \left(1+z_{\rm EdS}(t_\emi) \right) H_0 - H(z_{\rm EdS}(t_\emi)) \, , 
\end{align} 
and compute the difference between $\delta z_u$ and $z_{\rm EdS}^\text{alternative}$ 
at equal values of $\lambda_\emi$ (or $t_\emi$) instead of equal values of redshift. 
\newline\indent
Note also that the steep edges of the shaded area in this as well as other figures are not actually vertical, but merely very steep, corresponding to the steep density profile of the studied model.


\section{Conclusion} 
\label{sec:conclusion} 
We considered a formalism for computing the redshift drift in a general spacetime with an arbitrary observer and arbitrary emitting sources, and investigated the special case of an 
LTB model. We pointed out the importance of the position drift of the photons arriving at the observer, and that different values of position drift correspond to different potential emitters of the signal. 
For the LTB model we find that the contributions to the redshift drift from terms involving the drift of the viewing angle of light are several orders of magnitude smaller than the dominant Ricci and Weyl contributions. We can therefore to a good approximation neglect these complicating factors. 
Since we find that the drift of the viewing angle almost vanishes
once the ray has traversed the LTB structure, cumulative effects must be small. 
Hence, we expect the redshift drift signal to be dominated by its Ricci and Weyl contributions, 
also in Swiss cheese models based on LTB structures, but defer a detailed study of this point to upcoming work. Based on the similarities regarding the redshift behaviors in LTB and Szekeres models as studied in, e.g., section IV A in~\cite{Koksbang:2017arw}, we also expect the result to hold for  quasi-spherical Szekeres models and the corresponding Swiss-cheese models. 
\newline\indent
We considered the mean redshift drift signal for 1400 light rays with random impact parameters relative to the LTB structure. The mean redshift drift is dominated by the Ricci contribution, but we note that the mean Weyl contribution has an importance of around 5 \%, even in the FLRW region after the light rays have traversed the structure. We also note that inside the inhomogeneous region, the redshift drift associated with typical emitters deviates with several tens of percent from the ``background'' FLRW value. Although we expect such deviations to become less pronounced when light travels over greater distances, this indicates that the local effect of structures on the redshift drift signal may need attention when dealing with upcoming real data.

\vspace{6pt} 
\begin{acknowledgments}
We wish to thank Miko\l{}aj~Korzy\'nski for valuable comments. 
AH acknowledges funding from the European Research Council (ERC) under the European Union's Horizon 2020 research and innovation programme (grant agreement ERC advanced grant 740021--ARTHUS, PI: Thomas Buchert).  SMK is funded by the Carlsberg Foundation. The open source computer algebra system Maxima was used to verify computations of Christoffel symbols and Riemann tensors for the LTB metric. Some of the computations done for this project were performed on the UCloud interactive HPC system, which is managed by the eScience Center at the University of Southern Denmark.
\end{acknowledgments}

\appendix
\section{Multipole components for the LTB model}
\label{sec:LTBmultipole}
In this appendix, we provide a list with the explicit multipole components of equation \eqref{PicoefLTB} for the LTB model together with the explicit components of the drift of the viewing angle, $\bm{\kappa}$.
\newline\newline
For a random light ray, the components of $\bm{\kappa}$ are for the LTB metric given by
\begin{widetext}
\begin{align}
\begin{split}
{\kappa}^t &= 0\\
{\kappa}^r &= -\frac{1}{k^t}\left[ \left( 1-R\frac{\left( k^r\right) ^2}{\left( k^t\right) ^2} \right)\left(k^r_{,t} +\frac{R_{,t}}{2R}k^r \right) -\frac{k^rk^{\theta}}{\left( k^t\right) ^2}A^2\left( k^\theta_{,t} + \frac{A_{,t}}{A}k^\theta\right) -\frac{k^rk^\phi}{\left( k^t\right) ^2}A^2\sin^2(\theta)\left(k^{\phi}_{,t} +\frac{A_{,t}}{A}k^\phi \right)     \right] \\
{\kappa}^\theta & = -\frac{1}{k^t}\left[ -\frac{k^\theta k^r}{(k^t)^2}R\left(k^r_{,t} + \frac{R_{,t}}{2R}k^r \right) + \left(1-\frac{(k^{\theta})^2}{(k^t)^2}A^2 \right)\left(k^\theta_{,t} + \frac{A_{,t}}{A}k^{\theta} \right) -\frac{k^{\theta}k^\phi}{(k^t)^2}A^2\sin^2(\theta)\left(k^\phi_{,t} + \frac{A_{,t}}{A}k^\phi \right)   \right]\\
{\kappa}^{\phi}& = -\frac{1}{k^t}\left[  -\frac{k^\phi k^r}{(k^t)^2}R\left( k^r_{,t} +\frac{R_{,t}}{2R} \right) -\frac{k^\phi k^\theta}{(k^t)^2}A^2\left( k^\theta_{,t} +\frac{A_{,t}}{A}k^\theta\right) +\left(1-\frac{\left( k^{\phi}\right) ^2}{(k^t)^2}A^2\sin^2(\theta) \right)\left(k^\phi_{,t} + \frac{A_{,t}}{A}k^\phi \right)       \right] ,
\end{split}
\end{align}
\end{widetext}
from which the multipole term $-\kappa^\mu \kappa_\mu$ follows trivially.
\newline\indent
Using the Einstein equation, the monopole term of the redshift drift can simply be written as
\bea
\Sigma^{\it{o}}=	-\frac{1}{3}u^{\mu}u^{\nu} \mathcal{R}_{\mu\nu} = -\frac{4\pi G}{3}\rho.
\eea
This is the dominant redshift contribution for the considered setup. The other significant term is given by
\begin{align}
\begin{split}
\Sigma^{\bm{ee}}_{\mu \nu}e^{\mu}e^{\nu} &=      -  u^\rho u^\sigma  C_{\rho \mu \sigma \nu}e^{\mu}e^{\nu}  \\ &= -\left( C_{trtr}e^re^r + C_{t\theta t\theta}e^{\theta}e^{\theta} + C_{t\phi t\phi}e^{\phi}e^{\phi} \right) ,
\end{split}
\end{align} 
with 
\bea
&&C_{trtr} = -\frac{1}{2}R_{,tt} + \frac{1}{4}\frac{R_{,t}^2}{R} -\frac{4}{3}\pi G \rho R\\
&&C_{t\theta t\theta}= -AA_{,tt}-\frac{4}{3}\pi G\rho A^2\\
&&C_{t\phi t\phi} = -AA_{,tt}\sin^2(\theta)-\frac{4}{3}\pi G\rho A^2\sin^2(\theta).
\eea 
Finally, the shear term, which turns out to be negligible in the considered setup, is given by
\begin{align}
\begin{split}
\Sigma^{\bm{e\kappa}}_{\mu \nu}e^\mu \kappa^\nu &=  2 \sigma_{\mu \nu} e^\mu \kappa^\nu  \\ & = 	\frac{2}{3}\Sigma \left(-2Re^r \kappa^r + A^2 e^\theta \kappa^\theta + \sin^2(\theta)A^2 e^\phi \kappa^\phi \right) ,
\end{split}
\end{align}
where $\Sigma\equiv\frac{A_{,t}}{A} - \frac{A_{,tr}}{A_{,r}}$.

\section{Geodesic deviation and drift effects}
\label{sec:GD} 
In the study of drift effects we are interested in following the same emitter over time, and to consider the temporal change in various observable signals associated with that emitter. 
In order to describe drift signals in mathematical detail, we thus need to define a connecting congruence of photons between the observer worldline and the emitters of consideration. 
There are two ways that we might approach the selection of an emitter as viewed from a given observer worldline: i) we might simply consider \emph{a priori} fixing the emitter worldline. This uniquely determines a connecting congruence of null rays in the absence of caustics; or
ii) we can consider a fixed congruence of photons intersecting the observer worldline and deduce the class of potential emitters that intersect this congruence with their worldlines. From this class we might further identify a unique emitter from an appropriate criterion.  

There can be advantages of both approaches. For the purpose of explicit calculation, it can in practice be computationally difficult to construct the connecting null congruence between the observer and the fixed emitter as in the first approach. 
Thus, it is sometimes more convenient to take the second approach and simply consider the emitters that happen to intersect a given null congruence as initialised at the observer.  
We shall describe the latter approach here. See \cite{Korzynski:2017nas} for details on the first approach. 

Let the observer of interested be represented by its worldline $\gamma_o$ as generated by the 4-velocity $\bm{u}_o$. 
We consider a central null geodesic as received at the point of observation $\obs$ on $\gamma_o$, and we further consider a bundle of null geodesics around this central null ray that form a non-intersecting congruence. We might consider an appropriate extension of the congruence along the observer worldline to form a two-dimensional congruence. We might also consider a small extension of the congruence in the space orthogonal to $\bm{u}_\obs$ and the central incoming null ray in order to form a four-dimensional congruence of null rays. 
In any of the cases, there will be a 1-parameter family of null geodesics $\Gamma_{o}$ with 4-momentum field $\bm{k}$ intersecting the observer's worldline $\gamma_o$. 
We might ask which emitters that could have sent this family of photons that were later received by the observer. 
For a source to have emitted the null geodesics in $\Gamma_{o}$ its wordline must intersect thefamily of null lines of $\Gamma_{o}$. 
Formally, this is equivalent to demanding that the emitter 4-velocity is a deviation vector of $\Gamma_{o}$. 
Thus, assuming that a source wordline $\gamma_e$ intersects the central null ray at a point $\emi$; for this source to be emitter of $\Gamma_{o}$, we require that its 4-velocity satisfies  
$u^\mu_\emi = (E_\emi/E_\obs )X^\mu_\emi$, with $E= - k^\mu u_\mu$, and where $X^\mu$ is given by the propagation law 
\bea
\label{Liedeviation}
k^\nu \nabla_\nu X^\mu -  X^\nu \nabla_\nu k^\mu = \mu k^\mu  \, , \qquad X^\mu_\obs = u^\mu_\obs \, , 
\eea 
where $\mu$ is an arbitrary function, only restricted by the requirement that $X^\mu$ remains time-like, representing the possible parameterizations of the rays (with affine parameterizations characterised by $k^\nu \nabla_\nu \mu = 0$). 
The choice of proportionality constant in $u^\mu_\emi = (E_\emi/E_\obs )X^\mu_\emi$ is compatible with the conservation law $X^\mu k_\mu = (X^\mu k_\mu)_\obs = - E_\obs$ following from \eqref{Liedeviation}. 
The solutions to \eqref{Liedeviation} can be reformulated as 
\bea
\label{devdec}
X^\mu = \tilde{X}^\mu  + \alpha k^\mu \, , \qquad k^\mu\nabla_\mu \alpha = \mu  \, , \quad  \alpha_\obs = 0 \, , 
\eea 
with $\tilde{X}^\mu$ given by the solution to the propagation law without source term 
\bea
\label{Liedeviation2}
k^\nu \nabla_\nu  \tilde{X}^\mu -   \tilde{X}^\nu \nabla_\nu k^\mu = 0  \, , \qquad \tilde{X}^\mu_\obs = u^\mu_\obs \, , 
\eea 
such that $\tilde{X}^\mu$ obeys the usual geodesic deviation equation
\bea
\label{GDE}
k^\alpha \nabla_\alpha (k^\beta \nabla_\beta \tilde{X}^\mu) =  R^{\mu}_{ \, \alpha  \beta \nu} k^\alpha k^\beta \tilde{X}^\nu \, , 
\eea 
with a unique solution from the initial conditions $\tilde{X}^\mu_\obs = u^\mu_\obs$ and $\tilde{X}^\beta \nabla_\beta k^\mu \rvert_{\obs}$.  
The condition $\alpha \! >\! \tilde{X}^\mu \tilde{X}_\mu/E_\obs/2$ ensures that $X^\mu$ is timelike. 
The class of possible emitters of the null congruence $\Gamma_{o}$ as received by the observer are described by the class of tangent vectors given by \eqref{devdec} and satisfying the time-like condition. 
Conversely, emitters with 4-velocities that are not proportional to any of the tangent vectors in the class  \eqref{devdec} could \emph{not} have emitted the photons of $\Gamma_o$, and describing the drift effects of such emitters thus requires considering other appropriate photon congruences. 

In practice, for a given photon congruence, we might solve for the possible emitters of the photons intersecting the observer worldline by first solving (\ref{Liedeviation2}) and then considering the allowed class of transformations of the emitter tangent vector \eqref{devdec}. 
At each point along the central null ray, $\alpha$ might be chosen to uniquely determine an emitter 4-velocity $n^\mu = X^\mu/ (-X^\nu X_\nu)^{\frac{1}{2}}$. 
For instance, $\alpha$ might be chosen in a way that maximises $n^\mu U_\mu $ for a given preferred 4-velocity $\bm{U}$, which might not itself be intersecting $\Gamma_o$.  

We note that for a given emitter associated with a solution $X^\mu$ to \eqref{Liedeviation}, the position drift of the emitter on the observers sky is 
\bea
\label{pdrift}
\kappa_\obs^\mu &\equiv& p^{\mu}_{\, \nu} u^\alpha \nabla_\alpha e^\nu \rvert_{\obs}  = - p^{\mu}_{\, \nu}  \frac{1}{E} u^\alpha \nabla_\alpha k^\nu \rvert_{\obs}  + p^{\mu}_{\, \nu}  a^\nu \rvert_{\obs} \nonumber  \\ 
&=&- p^{\mu}_{\, \nu}  \frac{1}{E} k^\alpha \nabla_\alpha X^\nu \rvert_{\obs}  + p^{\mu}_{\, \nu}  a^\nu \rvert_{\obs} \nonumber  \\  
&=& - p^{\mu}_{\, \nu}  \frac{1}{E} k^\alpha \nabla_\alpha \tilde{X}^\nu \rvert_{\obs}  + p^{\mu}_{\, \nu}  a^\nu \rvert_{\obs} \, . 
\eea 
The last equality shows that the position drift is invariant under transformations of the source's tangentvector of the type \eqref{devdec}, and follows from \eqref{devdec} and the orthogonality between $k^\mu$ and the screen space projector $p^{\mu}_{\, \nu} \equiv - \frac{k^{\mu}}{E} \frac{k_{\nu}}{E} + \frac{k^{\mu}}{E}  u_\nu +   u^\mu  \frac{k_{\nu}}{E} + g^{\mu}_{\, \nu} $ as defined on the observer worldline. 
It follows that the position drift is determined by the initial conditions $u^\mu_\obs$, $a^\mu_\obs$,  $k^\mu_{\obs}$ and $u^\alpha \nabla_\alpha k^\mu \rvert_{\obs} \! =\! k^\alpha \nabla_\alpha \tilde{X}^\mu \rvert_{\obs}$. 
Physically, the observed angular drift of the source is independent on the exact points of emission along $\Gamma_{o}$, and the position drift signal is given entirely from the initialisation of the congruence of null rays at the observer position. 

For a given emitter 4-velocity $n^\mu \equiv X^\mu/ (-X^\nu X_\nu)^{\frac{1}{2}}$ and associated photon energy $E_n \equiv - n^\mu k_\mu$, the redshift drift signal is 
\bea
\label{redshiftdrift}
\hspace*{-0.1cm} \frac{d z}{d \tau_o} \Bigr\rvert_{\obs} &&=    - \frac{E_n \rvert_\emi }{E_\obs} \frac{ u_o^\mu \nabla_\mu E }{E}  \Bigr\rvert_{\obs}  +  \frac{ n^\mu \nabla_\mu E_n }{E_n}  \Bigr\rvert_{\emi} \nonumber  \\     
\hspace*{-0.1cm}  &&=   - \frac{ n^\mu k_\mu \rvert_\emi }{E_\obs} \frac{  (a^\mu_o k_\mu + u_o^\mu u_o^\nu \nabla_{\mu} k_\nu ) \rvert_{\obs} }{E_\obs }   \nonumber  \\   && \quad \, + \, \frac{ a^\mu k_\mu  + n^\mu n^\nu \nabla_{\mu} k_\nu }{n^\mu k_\mu }  \Bigr\rvert_{\emi}  \, , 
\eea  
with emitter and observer 4-accelerations given by $a^\mu \equiv n^\nu \nabla_\nu n^\mu$ and $a_o^\mu  \equiv u_o^\nu \nabla_\nu u_o^\mu$. 
The emitter 4-acceleration is not constrained by the above geodesic deviation analysis, and must be chosen independently. 
We shall usually be interested in setting the 4-accelerations to zero, corresponding to the case of physical emitters and observers that are subject only to gravitational physics. 
We can exploit that 
\bea
\label{trans}
\hspace*{-0.6cm} \frac{n^\mu n^\nu \nabla_{\mu} k_\nu }{(n^\mu k_\mu)^2 } \Bigr\rvert_{\emi} = \frac{1}{E^2_\obs} X^\mu X^\nu \nabla_{\mu} k_\nu = \frac{1}{E^2_\obs} \tilde{X}^\mu \tilde{X}^\nu \nabla_{\mu} k_\nu 
\eea   
to rewrite \eqref{redshiftdrift} as 
\bea
\label{redshiftdrift2}
\hspace*{-0.5cm} \frac{d z}{d \tau_o} \Bigr\rvert_{\obs} &=&      - \frac{ n^\mu k_\mu \rvert_\emi }{E_\obs} \left[ \frac{  (a^\mu_o k_\mu + u_o^\mu u_o^\nu \nabla_{\mu} k_\nu ) \rvert_{\obs} }{E_\obs } \right. \nonumber  \\ 
&&  \left. - \frac{ (E^2_\obs a^\mu k_\mu/(n^\mu k_\mu)^2  + \tilde{X}^\mu \tilde{X}^\nu \nabla_{\mu} k_\nu )\rvert_{\emi} }{E_\obs }   \right]  \! . 
\eea  
Thus the final redshift drift signal depends only on the components of $\nabla_{\mu} k_\nu$ as projected onto the canonical deviation vector $\tilde{X}^\mu$. 
The evolution of the velocity vector $\tilde{X}^\mu \nabla_{\mu} k_\nu$ can in turn be calculated along the central null ray once from the geodesic deviation equation \eqref{GDE}, where the right hand side is known once $\tilde{X}^\mu$ has been determined from the initial conditions $\tilde{X}^\mu_\obs = u^\mu_\obs$ and $u^\alpha \nabla_\alpha k^\mu \rvert_{\obs} $. 
Thus, the final expression for redshift drift depends only on the initial conditions for $u^\mu_\obs$, $a^\mu_\obs$,  $k^\mu_{\obs}$ and $u^\alpha \nabla_\alpha k^\mu \rvert_{\obs}$ together with $a_\emi^\mu$ and the transforming parameter $\alpha_\emi$. The latter parameter determines the photon energy as evaluated at the emitter $E_n \rvert_\emi = -n^\mu k_\mu \rvert_{\emi}$. 

The expression \eqref{redshiftdrift2} makes explicit that the redshift drift signal depends only on the extension of $k^\mu$ on the observers worldline through $u^\alpha \nabla_\alpha k^\mu \rvert_{\obs}$, and does \emph{not} depend on the initialisation of any of the other independent components\footnote{The components of $\nabla_\mu k_\nu \rvert_{\obs}$ are constrained by the null requirement $k^\nu \nabla_\mu k_\nu = 0$ and the geodesic requirement $k^\mu \nabla_\mu k^\nu = 0$.} of $\nabla_\mu k_\nu$. Any intermediate calculation making use of these should thus cancel for the final redshift drift signal. 




\bibliography{refs}

\end{document}